\documentclass{ajour}
\usepackage{epsf}
\newcommand{\bs}{\hskip -3pt} 

\begin{document}

\title{Finite size scaling and the role of the thermodynamic ensemble
  in the transition temperature of a dilute Bose gas.  } \author{Erich
  J. Mueller, Gordon Baym, and Markus Holzmann} \affil{Department of
  Physics, University of Illinois at Urbana-Champaign,\\ 
  1110 W Green St. Urbana IL, 61801.}
\date{\today}
\titlerunninghead{Finite size scaling\ldots} \authorrunninghead{EJ
  Mueller, G Baym, and M Holzmann}

\abstract{ We study the Bose-Einstein condensation phase transition in
  a weakly interacting gas through a perturbative analysis of finite
  systems.  In both the grand canonical and the canonical ensembles,
  perturbation theory suffers from infrared divergences and cannot
  directly determine the transition temperature in the thermodynamic
  limit.  However, in conjunction with finite size scaling,
  perturbation theory provides a powerful calculation tool.  We
  implement it here to estimate a shift in the transition temperature
  in the canonical ensemble consistent with grand canonical
  calculations.  } 
\begin{article}
  
  Wilkens, Illuminati, and Kr\"amer \cite{wilkens} derive a surprising
  and extremely interesting result, namely that as a function of
  interaction strength the transition temperature of a dilute Bose gas
  behaves differently in the canonical and grand canonical ensembles.
  They conclude that in the canonical ensemble, the transition
  temperature {\em decreases} with increasing interaction strength,
  whereas the opposite behavior has been established in calculations
  in the grand canonical ensemble \cite{gce,bbz}.  The latter agree
  with numerical calculations in the canonical ensemble which
  extrapolate to the thermodynamic limit using finite size scaling
  techniques \cite{gruter,holz}.
  
  In this paper we reformulate the ideas of Wilkens et al.
  \cite{wilkens} to provide a more explicit comparison with its grand
  canonical counterparts and with numerical calculations.  The
  calculation is perturbative in nature; we show that such a
  perturbative scheme fails in the thermodynamic limit due to the
  presence of infrared divergences.  As in the grand canonical
  ensemble, long distance properties dominate thermodynamic quantities
  of the system, as expected for a second order phase transition.
  Standard finite size scaling, as used in \cite{holz}, provides a
  tool to overcome these difficulties, giving a precise scheme to
  extrapolate to the thermodynamic limit.
  
  This paper is structured as follows.  First, to lowest order in the
  interaction strength, we derive a perturbative expression for the
  transition temperature.  Next, by considering higher order terms, we
  show that the perturbation series diverges in the thermodynamic
  limit.  Finally, we use finite size scaling in conjunction with the
  perturbation series to correctly derive the shift in the transition
  temperature.  Our first section closely follows \cite{wilkens}, as
  one of our primary goals is to reconcile their calculation with
  conventional results.
  
  Consider a system of bosons interacting through a short range
  potential described by the Hamiltonian
\begin{equation}
H=\sum_q \frac{\hbar^2 q^2}{2m} b_q^\dagger b_q +H_{\rm int},
\end{equation}
where
\begin{equation}
H_{\rm int}=
\frac{2\pi\hbar^2 a}{mV}\sum_{\bf pkq} b_p^\dagger b_q^\dagger b_{q-k}b_{p+k},
\end{equation}
and $a$ is the scattering length, $b_q$ is the annihilation operator
for particles with momentum $q$, $m$ is the particle mass, and
$V\equiv L^3$ is the volume of the system.  The free energy to first
order in $a$ is
\begin{equation}\label{f1}
F(a)=F_0+\langle H_{\rm int}\rangle,
\end{equation}
where the expectation value is in the {\em free} ensemble, and
$F_0\equiv F(a=0)$ is the free energy of the non-interacting system.
The basic technical difference between the canonical and grand
canonical ensembles arises in factoring the four point expectation
value in Eq.~(\ref{f1}); in the grand canonical ensemble
\begin{eqnarray}
\sum_{pqk}\langle b_p^\dagger b_q^\dagger
b_{q-k}^{\phantom\dagger}b_{p+k}^{\phantom\dagger}\rangle_{\rm GC}
&=& \sum_{pqk}
\left(\langle b_p^\dagger b_{q-k}^{\phantom\dagger}\rangle\langle
b^\dagger_{q}b_{p+k}^{\phantom\dagger}\rangle+
\langle b_p^\dagger b_{p+k}^{\phantom\dagger}\rangle\langle
b_{q}^\dagger b_{q-k}^{\phantom\dagger}\rangle\right)\\
&=&2\langle N\rangle^2,
\end{eqnarray}\bs
where $\langle N\rangle=\sum_k\langle b_k^\dagger b_k\rangle$ is the
mean number of particles.  In the canonical ensemble
\begin{equation}\label{fprime}
\sum_{pqk}
\langle b_p^\dagger b_q^\dagger
b_{q-k}^{\phantom\dagger}b_{p+k}^{\phantom\dagger}\rangle_{\rm C}=
2N^2-\sum_{p} \langle N_p(N_p+1)\rangle,
\end{equation}
where $N_p=b_p^\dagger b_p$.  Thus the first order shift in free
energy differs in the two ensembles by the term
\begin{equation}\label{diff}
\langle H_{\rm int}\rangle_{\rm C} -\langle H_{\rm int}\rangle_{\rm GC}
= -\frac{2\pi\hbar^2 a}{mV}\sum_{ p}\langle N_p (N_p+1)\rangle.
\end{equation}
Although the coefficient of this sum is of order $1/V$, the sum itself
diverges as $V^{4/3}$ at the Bose condensation transition.  This term
of order $V^{1/3}$ is the source of the shift in the critical
temperature described in \cite{wilkens}.  The contribution
proportional to $N^2$ can not affect the critical temperature, since
this term just adds a constant to the free energy.  We now restrict
ourselves to the canonical ensemble to determine the role of the term
(\ref{diff}). In all subsequent manipulations both the number of
particles $N$ and the system size $V$ are held fixed.

The order parameter for the Bose-Einstein condensation transition is
$N_0$, the number of condensed particles.  To explore the transition
we study the probability of having $N_0$ particles in the condensate.
This distribution function, $P(N_0)$, is
\begin{equation}
P(N_0) = \frac{1}{Z_N}\mathop{{\rm Tr}}_{N,N_0\, {\rm fixed}}e^{-\beta H}
=\frac{1}{Z_N}e^{-\beta F(N,N_0)}, 
\end{equation}
where the trace is taken at fixed $N$ and $N_0$, and $\beta$ is the
inverse temperature.  This equation defines the free energy
$F(N,N_0)$, and uses the partition function, $Z_N=\sum_{N_0}e^{-\beta
  F(N,N_0)}$, to normalize the probability distribution.  The high
temperature normal phase of the system is characterized by a
monotonically decreasing $P(N_0)$, while the low temperature condensed
phase has $P(N_0)$ peaked at $N_0\neq0$.  At sufficiently high
temperatures, $N_0$ is Gibbs distributed, $P(N_0)\propto e^{\beta\mu
  N_0}$, where $\mu=\partial F/\partial N<0$ is the chemical
potential.  At zero temperature in the non-interacting gas all the
particles are condensed and $P(N_0)=\delta_{N_0,N}$.  At some
intermediate temperature the distribution becomes flat at $N_0=0$.
Wilkens et al. define $T_c$ by extrapolating this crossover
temperature to the thermodynamic limit.

In terms of $F(N,N_0)$, Wilkens et al.'s criterion for $T_c$ gives an
implicit equation for the critical temperature of the interacting
system $T_c^{(a)}$,
\begin{equation}\label{df}
\left.
\frac{\partial F(N,N_0)}{\partial N_0}\right|_{N_0=0,T=T_c^{(a)}}
=0.
\end{equation} 
As in Fermi liquid theory, $\partial F/\partial N_0$ is the energy of
a $k=0$ quasiparticle measured from the chemical potential \cite{flt},
and can therefore be expressed as
\begin{equation}
\frac{\partial F}{\partial N_0}=\Sigma(k=0,\omega=0)-\mu,
\end{equation}
where $\Sigma(k,\omega)$ is the self-energy at momentum $k$ and energy
$\omega$.  Thus this criterion for the critical temperature is
essentially that used by Baym et al. \cite{bbz} in the grand canonical
ensemble.  An important difference between the two approaches is that
in the present calculation only quantities at $N_0=0$ are involved.
In the canonical ensemble the fluctuations in $N_0$ are very large at
the critical temperature, $\langle N_0^2\rangle-\langle N_0\rangle ^2
\sim N^{4/3}$ \cite{ill}; as we shall see, the criterion (\ref{df})
yields a qualitatively different shift in the transition temperature
if the derivative is evaluated at the expectation value of $N_0$
rather than at $N_0=0$.

We now expand the criterion (\ref{df}) for $T_c$ in powers of $a$, to
calculate perturbatively the transition temperature, $T_c^{(a)}$, of
the interacting system.  Since $\partial F/\partial N_0$ is evaluated
at $T=T_c^{(a)}$, we must consider not only the explicit variation of
$F$ with $a$, but also the implicit contribution due to the dependence
of $T$ on $a$.  We use the decomposition $F(a)=F_0+\Delta F(a)$, where
$F_0$ is the free energy of the non-interacting gas and $\Delta F(a)$
is the correction due to interactions.  In the free system, the
condensate only contributes to the free energy by reducing the
occupation of other modes, i.e.,
\begin{equation}
\frac{\partial F_0(N,N_0)}{\partial N_0}=-
\frac{\partial F_0(N,N_0)}{\partial N}
\equiv-\mu_0, 
\end{equation}
which defines the free chemical potential $\mu_0$, a function of $N$,
$N_0$ and $T$.  By construction, when $N_0=0$ this chemical potential
vanishes at the transition temperature of the non-interacting gas,
$T_c^{(0)}$, and to first order in the interaction,
\begin{eqnarray}
\left.\frac{\partial F_0}{\partial N_0}\right|_{N_0=0,T=T_c^{(a)}}&=&
-\mu_0(N_0=0,T=T_c^{(a)})\\
&=&
-\Delta T_c \left.
\frac{\partial\mu_0}{\partial
  T}\right|_{N_0=0,T=T_c^{(0)}}
+{\cal O}(a^2).  
\end{eqnarray}\bs%
The derivative is taken at fixed $N$ and $N_0$, and $\Delta
T_c=T_c^{(a)}-T_c^{(0)}$ is the shift in the transition temperature
for scattering length $a$.  Thus, to first order in $a$, the left hand
side of Eq.~(\ref{df}) becomes $\partial (\Delta F)/\partial
N_0-\Delta T_c\, \partial \mu_0/\partial T$, evaluated at
$T=T_c^{(0)}$ and $N_0=0$.  Solving for $\Delta T_c$, we have
\begin{equation}\label{pert}
\Delta T_c = \left.\frac{(\partial (\Delta F)/\partial
  N_0)}{(\partial\mu_0/\partial T)}\right|_{N_0=0,T=T_c^{(0)}}.
\end{equation}
Aside from the use of continuous derivatives in place of Wilkens et
al.'s discrete derivatives, this is the result of \cite{wilkens}.
Correctly evaluating these functions for a finite sized system is
challenging.  We estimate their magnitude by replacing the canonical
expectation values in Eq.~(\ref{fprime}) by the grand canonical result
$\langle N_k (N_k+1)\rangle = 2 \langle N_k\rangle(\langle
N_k\rangle+1)$, and approximately writing $\langle N_k\rangle \approx
(e^{\beta(\epsilon_k-\mu_0)}-1)^{-1}$.  This assumption provides a
simple relationship between $N=N_0+\sum_k\langle N_k\rangle$ and
$\mu_0$. Introducing the number of excited particles $N_{\rm
  ex}=N-N_0$, we may write
\begin{eqnarray}
\frac{\partial (\Delta F)}{\partial N_0}
&=& -\frac{\partial (\Delta F)}{\partial N_{\rm ex}}
= -\left(\frac{\partial (\Delta F)}{\partial \mu_0}\right)
\left(\frac{\partial \mu_0}{\partial N_{\rm ex}}\right)\\
&=& \left(\frac{\partial (\Delta F)}{\partial \mu_0}\right)
\left(\frac{\partial \mu_0}{\partial T}\right)_{N_{\rm ex}}
\left(\frac{\partial T}{\partial N_{\rm ex}}\right)_{\mu_0}.
\end{eqnarray}\bs%
Since all quantities are evaluated in the free ensemble, the
derivatives are straightforwardly evaluated, leading to
\begin{equation}\label{shift}
\frac{\Delta T_c}{T_c} \approx -\frac{8\pi\hbar^2 a}{3mNV k_B T}
\sum_{k\neq 0} \langle N_k\rangle^3,
\end{equation}
where $\langle N_k\rangle$ is evaluated at $\mu_0=0$.  The sum is
infrared divergent, scaling as $V^{2}$, and yielding a finite
temperature shift proportional to $-a n^{1/3}$, where $n=N/V$.  The
constant of proportionality is of the same order of magnitude as the
one calculated in \cite{wilkens} using a sophisticated series of
asymptotic expansions; its exact numerical value is unimportant here.
The key observation is that contrary to the expected behavior, the
temperature shift predicted by this argument is negative.

This negative temperature shift depends crucially upon the constraint
$N_0=0$.  At finite $N_0$, the numerator of Eq.~(\ref{pert}) has an
additional contribution due to the derivative of the $p=0$ term of the
sum in Eq.~(\ref{diff}).  This contribution has the opposite sign, and
dominates when $N_0\sim N^{2/3}$, yielding a positive temperature
shift.  As already emphasized, at the critical temperature, the
expectation value of $N_0$ is of order $N^{2/3}$.

We now explore the validity of this perturbation expansion,
demonstrating that it breaks down in the thermodynamic limit.  Higher
order terms in the expansion of the free energy (\ref{f1}) involve
higher powers of the interaction $H_{int}$. As in first order, the
most divergent terms occur when all of the momenta are equal and, at
$T_c$, these terms are of relative size $\langle (\beta
H_{int})^{m}\rangle/ \langle (\beta H_{int})^{m-1}\rangle\sim
aL/\lambda^2$, where $L = V^{1/3}$ is the length of the system, and
$\lambda = (2\pi\hbar^2/ mk_b T)^{1/2}\sim n^{-1/3}$ is the thermal
wavelength, of order the interparticle spacing.  Thus we see that the
perturbation expansion is valid only for sufficiently small $\eta
\equiv aL/\lambda^2$.

In any finite system the transition temperature calculated above
corresponds to a crossover of finite width $\delta T$.  One can
estimate the width of the crossover from the fluctuations in the
number of condensed particles; since the latter scale as \cite{ill},
\begin{equation}\label{flc}
\delta T/T\sim\delta N_0/N \sim N^{-1/3}.
\end{equation}
As long as $\delta T\ll \Delta T_c$, the shift is well defined.  Since
$\Delta T \propto - an^{1/3}$, the ratio $\Delta T/\delta T$ is of
order $aL/\lambda^2 =\eta$, and the shift only is well defined for
$\eta\gg1$.  Thus in the limit of small $\eta$, where the expansion of
the free energy converges, the calculated change in $T_c$ is smaller
than the width of the transition and cannot be physically significant.
In the other limit $\eta\gg1$ the expansion of the free energy breaks
down.  Hence this calculation, as it stands, cannot tell us anything
about the transition temperature of a weakly interacting Bose gas.

Although perturbation theory breaks down in the the thermodynamic
limit, it can be used to learn the properties of small systems where
$L \ll \lambda^2/a$.  We now discuss how finite size scaling enables
one to learn about the $L=\infty$ phase transition by investigating
how physical quantities scale with $L$ in these small systems.  This
technique is commonly used in numerical simulations where it is not
feasible to simulate an infinite system.  The central assumption of
finite size scaling is that sufficiently close to the critical point
all physical quantities scale as functions of the ratio of the
correlation length $\xi$ to the system size $L$.  For example the
order parameter scales as
\begin{equation}\label{sc1}
\frac{\langle N_0\rangle}{V}\sim L^{-y} \Phi(L/\xi),
\end{equation}
where $y=\beta/\nu=1$ is the ratio of the critical exponents for
$N_0/V$ and the correlation length, and $\Phi$ is a scaling function.
As $L/\xi\to\infty$, this function must diverge as $\Phi(L/\xi)\sim
(L/\xi)^y$, while as $L/\xi\to 0$, $\Phi$ approaches a constant.  The
latter limit gives a systematic method for finding the critical point
($\xi\to\infty$) by looking solely at the properties of a finite
system.  In numerical calculations \cite{gruter,holz} one plots $L^y
\langle N_0\rangle/V$, or a related quantity such as the superfluid
density, as a function of temperature for different system sizes.
According to the scaling hypothesis all of these curves should
intersect at the critical temperature.

Relations similar to Eq.~(\ref{sc1}) also hold for higher moments of
the order parameter and imply that at $T_c$ the probability
distribution function can be written as
\begin{equation}\label{pscal}
P(N_0)= \frac{\lambda^2}{L^2} \psi(N_0\lambda^2/L^2),
\end{equation}
with some scaling function $\psi$.  We calculate the critical
temperature of the interacting system by finding the temperature at
which $P(N_0)$ has this scaling form.

\begin{figure}[!tpb]
  \epsfxsize=\columnwidth \centerline{\epsfbox{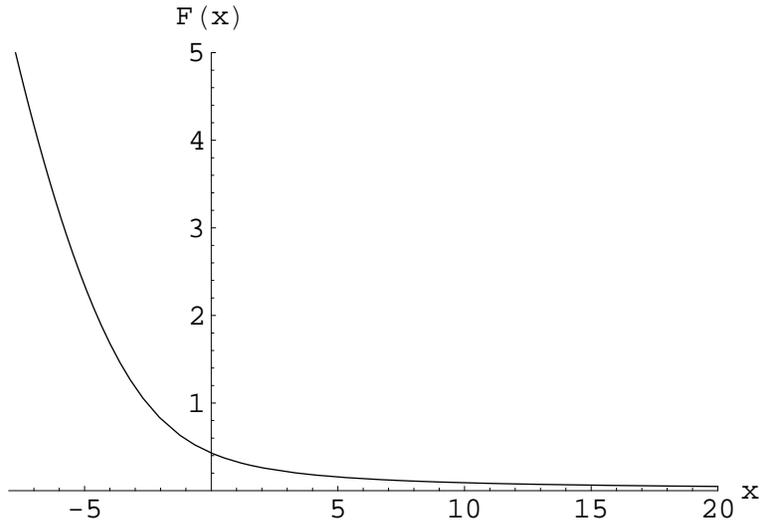}}
\caption{The function $F(x)$, defined as the solution to
  Eq.~(\ref{ffdef}), which relates the chemical potential and density
  of a non-interacting ideal bose gas via
  Eq.~(\ref{fuse}).}\label{figf}
\end{figure}

Before using this procedure to calculate $\Delta T_c$, we verify that
the scaling relations, Eqs.~(\ref{pscal}) and (\ref{sc1}), hold in the
non-interacting gas.  We first derive Eq.~(\ref{sc1}) in the grand
canonical ensemble where the argument is particularly simple.  The
general strategy is to fix the average density $n=N/V$ and the
temperature $T$, and look at how the order parameter $N_0$ varies with
the system size $L$.  To carry out this approach we need an expression
for the chemical potential $\mu$ as a function of $n$, $T$, and $L$,
which requires inverting the relationship,
\begin{eqnarray}
n &=& \frac{1}{V}\sum_k \frac{1}{e^{\beta(\epsilon_k-\mu)}-1}\\\label{dens}
&\approx& \frac{1}{\lambda^3} g_{3/2}(e^{\beta\mu})+N_0/V,
\end{eqnarray}\bs%
where $g_{3/2}(z)=\sum_j z^j/j^{3/2}$ is a polylogarithm function.
The inversion is performed by expanding (\ref{dens}) in powers of
$\beta\mu$, noting that $N_0\approx-1/\beta\mu$.  One finds
\begin{equation}
n = \frac{1}{\lambda^3} \left[\zeta(3/2)
+\frac{\lambda}{L}\left(-\frac{1}{\beta\mu L^2/\lambda^2}
-2\sqrt{-\pi\beta\mu L^2/\lambda^2}\right)+\cdots\right];
\end{equation}
for $\beta\mu\sim\lambda^2/L^2\sim 1$, the neglected terms are of
relative order $\lambda/L$.  The terms proportional to $1/\beta\mu$
and $\sqrt{-\beta\mu}$ are respectively the contributions from the
condensed and non-condensed particles.  Finding $\beta\mu$ as a
function of $n$, $T$, and $L$, requires solving a cubic equation.  We
define the function $F(x)$, plotted in Fig.~\ref{figf}, as the
solution to
\begin{equation}\label{ffdef}
\frac{1}{F(x)}-2\sqrt{\pi F(x)}-x=0,
\end{equation}
so that the chemical potential can be expressed as,
\begin{eqnarray}\label{fuse}
\beta\mu &=&-\frac{\lambda^2}{L^2} 
F\left(\frac{L}{\lambda}\left(\lambda^3 n -\zeta(3/2)\right)\right).
\end{eqnarray}\bs%
The positive, monotonic $F(x)$ has the properties
\begin{eqnarray}
F(0)&=&(4\pi)^{-1/3}\\
F(x)&
\displaystyle \mathop{{\longrightarrow}}_{x\to-\infty}
&x^2/4\pi\\
F(x)&
\displaystyle \mathop{{\longrightarrow}}_{x\to+\infty}
&1/x.
\end{eqnarray}\bs%
Thus, as $L\to\infty$, the order parameter $N_0\approx -1/\beta\mu$,
has three distinct behaviors, corresponding to non-condensed,
critical, and condensed regimes, depending on whether $n$ is less
than, equal to, or greater than $\zeta(3/2)/\lambda^3$.  In the
non-condensed regime, $N_0$ is microscopic, in the condensed regime,
$N_0$ is extensive, and at the critical point, $N_0$ scales as
Eq.~(\ref{sc1}), with the predicted exponent, $y=1$, {\em i.e.},
\begin{eqnarray}
\label{sca}
n<\zeta(3/2)/\lambda^3,&\quad&N_0\sim L^{0}\\\label{scalc}
n=\zeta(3/2)/\lambda^3,&\quad&N_0\sim L^{2}\\\label{scb}
n>\zeta(3/2)/\lambda^3,&\quad&N_0\sim L^{3}.
\end{eqnarray}\bs%
The scaling at $T_c$, Eq.~(\ref{scalc}), is consistent with
(\ref{flc}), since at the critical point the mean value of $N_0$ is of
the same order as the fluctuations $\delta N_0$.

\begin{figure}[!tpbh]
  \epsfxsize=\columnwidth \centerline{\epsfbox{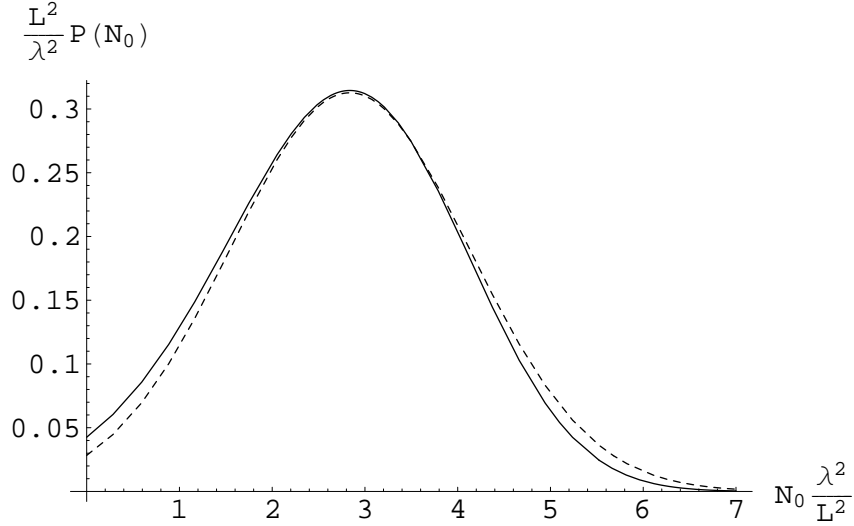}}
\caption{
  Probability distribution $P(N_0)$ for the number of condensed
  particles at the critical temperature in a non-interacting Bose gas
  within the canonical ensemble.  Dashed line is a Gaussian
  approximation.  At the critical point, $P$ has the scaling form
  (\ref{pscal}).  }\label{probdist}
\end{figure}

We now perform the equivalent calculation in the canonical ensemble.
Starting from the expression $N=N_0+\sum_k\langle N_k\rangle$, and
approximating $\langle
N_k\rangle=(e^{\beta(\epsilon_k-\mu_0(N_0))}-1)^{-1}$, we solve for
$\mu_0$ as a function of $N_0$.  Integrating $\mu_0(N_0)$ with respect
to $N_0$ yields $F(N,N_0)$, and $P(N_0)\propto e^{-\beta F(N,N_0)}$.
As $P(N_0)$ is peaked about the point where $\mu_0=0$, we can expand
$\sum_k \langle N_k\rangle$ in powers of $\mu_0$.  Standard
asymptotics \cite{ziff} yields a series which converges for
$|\beta\mu_0|<\pi\lambda^2/L^2$,
\begin{eqnarray}\label{nexs}
N_{\rm ex}&=&\sum_k\langle N_k\rangle
=\left(\frac{L}{\lambda}\right)^3\zeta(3/2)
+\left(\frac{L}{\lambda}\right)^2
f\left(\frac{\beta\mu L^2}{\lambda^2}\right) +{\cal
  O}(L/\lambda),\\\label{fdef}
f(x) &=& \sum_{k=0}^{\infty} \frac{x^k}{k!}C_3(k+1),
\end{eqnarray}\bs%
where the coefficients $C_d(k)$ are sums of the form,
\begin{equation}\label{coef}
C_d(k) =
\frac{\Gamma(k)}{\pi^k} \mathop{{\sum}^\prime}_{n_1,n_2,\ldots,n_d}
\frac{1}{\left(n_1^2+n_2^2+\cdots+n_d^2\right)^{k}};
\end{equation}
the prime denotes that the term $n_1=n_2=\cdots=n_d=0$ is omitted.
These constants are tabulated in Table~\ref{tab:cd}.
\begin{table}[!tpb]
\caption{
Values of the lattice sum $C_3(k)$, defined in Eq.~(\ref{coef}).}
\label{tab:cd}
\begin{tabular}{llllll}
\hline
k=1      &2      &3      &4&5&6      \\
\hline
-2.8374&1.6752&0.5419&0.4278&0.5039&0.7741\\
\hline
\end{tabular}
\end{table}
Inversion of the series gives
\begin{eqnarray}\label{mu}
\beta\mu
&=&\frac{\lambda^2}{L^2}
\left(-\frac{1}{C_3(2)}M-\frac{C_3(3)}{2 (C_3(2))^2}M^2 +\cdots\right),\\
M&=& N_0\frac{\lambda^2}{L^2}+C_3(1)-
\frac{\lambda^2}{L^2}\left(N-\frac{L^3}{\lambda^3}\zeta(3/2)\right),
\end{eqnarray}\bs%
from which the free energy is
\begin{eqnarray}\label{ff}
\beta F_0(N,N_0)&=&-\int\!\!dN_0\,(\beta\mu_0)\\\label{fb}
&=&\beta\bar F_0(N) + \frac{1}{2 C_3(2)}M^2 + \frac{C_3(3)}{6
  (C_3(2))^2}M^3+\cdots\\\label{fc}
&\equiv&\beta\bar F_0(N) + g(M).
\end{eqnarray}\bs%
where $\bar F_0(N)$ is an extensive function which is independent of
$N_0$, and $g(M)$ is defined by Eqs.~(\ref{fb},\ref{fc}).  If
$n=N/V=\zeta(3/2)/\lambda^3$, then $M$ depends on $L$ only through the
variable $N_0\lambda^2/L^2$, implying that $P(N_0)$ is of the form
(\ref{pscal}).  For any other value of the density, $M$ has additional
$L$ dependence, and $P(N_0)$ does not have the desired form, implying
that the scaling (\ref{pscal}) holds only at $T_c$.  In
Fig.~\ref{probdist} we plot $P(N_0)$ at the critical point.  The
Gaussian approximation, where only the term proportional to $M^2$ is
kept, is also plotted, and agrees quite well with the full result.

Having verified the scaling relationships for $P(N_0)$ in the
non-interacting gas, we now search for the critical temperature of the
interacting gas by perturbatively calculating $P(N_0)$ and finding the
temperature at which scaling holds.  We continue to use the
approximations that $\langle N_k (N_k+1)\rangle = 2 \langle
N_k\rangle(\langle N_k\rangle+1)$, and $\langle N_k\rangle \approx
(e^{\beta(\epsilon_k-\mu_0)}-1)^{-1}$.  To first order in $a$, at
temperature $T=T_c^{(0)}+\Delta T$, the free energy is,
\begin{eqnarray}
\beta F[N,N_0] &=& \beta \bar F_0 + g(M_0) 
-\frac{\Delta T}{T}\frac{\lambda^2}{L^2}N_0g^\prime(M_0)
\\\nonumber&&
+\frac{3}{2}\frac{\Delta T}{T}\frac{L}{\lambda}\zeta(3/2) g^\prime(M_0)
-\frac{2\pi\hbar^2 a}{mVT}\left(N_0^2+2\sum_{k\neq0}N_k(N_k+1)\right),
\end{eqnarray}\bs
where the argument $M_0= N_0(\lambda^2/L^2)+C_3(1)$ is the scaled
condensate number measured from the peak of the distribution.  The
first two terms are the free energy of the non-interacting gas at
$T_c^{(0)}$, while the remaining terms give the first order
corrections in $\Delta T$ and $a$.  These corrections are only small
if $\eta=aL/\lambda^2\ll1$.  The sum $\sum_{k\neq0} N_k(N_k+1)$ can be
identified with $\partial N_{\rm ex}/\partial (\beta\mu_0)$, and can
be expressed as a series in $\beta\mu_0L^2/\lambda^2$ via
Eq.~(\ref{nexs}).  Using Eq.~(\ref{mu}) to eliminate $\mu$, the
corrections are
\begin{eqnarray}
F(T)&=& F_0(T_c^{(0)})-\frac{\Delta T}{T}\frac{\lambda^2}{L^2}N_0
g^\prime(M_0) \\\nonumber&&
+\frac{L}{\lambda}
\left[
  \frac{3\zeta(3/2)}{2}
  \frac{\Delta T}{T}
  \left(
     \frac{1}{C_3(2)}M_0 +
     \frac{C_3(3)}{2C_3(2)^2}M_0^2+\cdots
  \right)
\right.\\\nonumber&&\left.
-\frac{2\pi\hbar^2a}{m\lambda^3 T}\left(
C_3(1)^2+2C_3(2)-
2\left(C_3(1)+\frac{C_3(3)}{C_3(2)}\right) M_0\right.\right.
\\\nonumber&&\quad\left.\left.
+\left(1+\frac{C_3(2)C_3(4)-C_3(3)^2}{C_3(2)^3}\right)
M_0^2+\cdots\right)
\right].
\end{eqnarray}\bs%
Comparing with Eq.~(\ref{pscal}), we see that
scaling holds if and only if the factor multiplied by $L/\lambda$
vanishes.  Eliminating the coefficient of the first power of $M_0$
enforces scaling near the peak of $P(N_0)$, in which case
\begin{eqnarray}
\frac{\Delta T}{T} &=& 
-\frac{8\pi\hbar^2 a}{3 m\lambda^3 T\zeta(3/2)} 
\left(C_3(1) C_3(2)+C_3(3)\right)\\
&\approx& 1.6 a n^{1/3}.
\end{eqnarray}\bs%
The coefficient $1.6$ should be compared with the numerical value
$2.3$ calculated by Holzmann and Krauth \cite{holz}.  The discrepancy
lies within the accuracy expected of our approximations.  The
important point to note is that the coefficient is positive and of
order unity.

The neglect of terms of higher order in $a$ during the calculation is
based on the assumption that they do not change the structure of the
scaling function.  (We note that recent calculations of $\phi^4$
theory on a lattice \cite{arnold,kash}, may indicate that this
assumption is not valid.)  A more involved study, where these higher
order terms are explicitly calculated would help verify whether
perturbation theory is valid within finite size scaling.

At this point it would be appealing to repeat the above calculation in
the grand canonical ensemble and explicitly verify that the two
ensembles yield the same shift in the transition temperature.  In the
grand canonical ensemble, first order perturbation theory changes the
energy of each momentum state by the same amount.  This shift can
therefore be absorbed into the chemical potential, leaving the
transition temperature unchanged.  The first effects start at higher
order; exploring how higher order perturbation theory in conjunction
with finite size scaling can be used to calculate the shift of $T_c$
in the grand canonical ensemble will be discussed in a future
publication.

In summary, we demonstrate that infrared divergences prevent the
direct application of perturbation theory to calculating the
transition temperature of a dilute Bose gas in the canonical ensemble.
We use scaling arguments to circumvent this problem and to evaluate
$\Delta T_c$ within the canonical ensemble, finding results which are
consistent with numerical calculations, and with analytic results
based on the grand canonical ensemble.

\begin{acknowledgment}
  This work was supported in part by NSF Grant PHY98-00978 and the
  NASA Microgravity Research Division, Fundamental Physics Program.
  We are grateful to Fabrizio Illuminati and Werner Krauth for seminal
  discussions.
\end{acknowledgment}

\end{article} 
\end{document}